\documentclass[twocolumn]{article} 

\usepackage{graphicx}
\usepackage{graphbox}
\usepackage{cite}
\usepackage{amssymb}
\usepackage{latexsym}
\usepackage{graphicx}
\usepackage{subfigure}
\usepackage{amsmath}
\usepackage{mathtools}
\usepackage{url}
\usepackage{pgf,tikz}
\usepackage{tikz}
\usetikzlibrary{spy}
\usepackage{xcolor}
\usepackage{algorithm}%
\usepackage{algorithmicx}%
\usepackage{algpseudocode}%
\usepackage{listings}%
\usepackage{graphbox}
\usepackage{tabularx}
\newcolumntype{L}{>{\raggedright\arraybackslash}X}
\graphicspath{ {images/} }

\begin{document}
\sloppy

\title{HPC-based Solvers of Minimisation Problems for\\ Signal Processing}%

\makeatletter
\onecolumn
{\fontsize{18pt}{20pt}\selectfont\bfseries\@title\par}
Simone Cammarasana
 \footnote{
\textbf{Simone Cammarasana}
CNR-IMATI, Via De Marini 6, Genova, Italy \\
simone.cammarasana@ge.imati.cnr.it
}
  , 
 Giuseppe Patan\`e
  \footnote{
 \textbf{Giuseppe Patan\`e} 
 CNR-IMATI, Via De Marini 6, Genova, Italy 
}

\makeatother

\begin{abstract}
Several physics and engineering applications involve the solution of a minimisation problem to compute an approximation of the input signal. Modern computing hardware and software apply high-performance computing to solve and considerably reduce the execution time. We compare and analyse different minimisation methods in terms of functional computation, convergence, execution time, and scalability properties, for the solution of two minimisation problems (i.e., approximation and denoising) with different constraints that involve computationally expensive operations. These problems are attractive due to their numerical and analytical properties, and our general analysis can be extended to most signal-processing problems. We perform our tests on the Cineca Marconi100 cluster, at the 26th position in the ``\emph{top500}'' list. Our experimental results show that PRAXIS is the best optimiser in terms of minima computation: the efficiency of the approximation is~$\textnormal{38\%}$ with 256 processes, while the denoising has~$\textnormal{46\%}$ with 32 processes.

\end{abstract}

%
\section{Introduction\label{SEC:INTRODUCTION}}
Minimisation methods are widespread for solving various physics and engineering problems. State-of-the-art optimisers account for several properties, such as global vs local search and exploitation of analytical derivatives (Sect.~\ref{SEC:RELATEDWORK}). Minimisation problems are also widely applied to signal processing for image analysis~\cite{huo2018towards}, 2D videos~\cite{cammarasana2022fast}, and graph signal processing~\cite{hendrickson2008graph}, with applications in biomedicine~\cite{gulo2019techniques}, astronomy~\cite{kocz2015digital}, and computer vision~\cite{webb1994high}. Solving signal processing requires a significant computational effort, and the real-time constraint~\cite{cammarasana2022real}\cite{le2019radar} requires to reduce the execution time.

The \emph{constrained optimisation by linear approximations} (COBYLA)~\cite{Powell1994} is applied for the simulation and prediction of structural and acoustic properties of a geometrical model~\cite{bos2002numerical}, image segmentation and classification~\cite{9996795}, and design of pressure vessel in terms of geometrical properties to minimise material and fabrication cost~\cite{woldemichael2016optimization}. The \emph{Limited-memory Broyden, Fletcher, Goldfarb, Shanno} (L-BFGS)~\cite{zhu1997algorithm} is applied to large-scale prediction problems on biomedical images co-registration~\cite{yang2015non}, protein structure~\cite{jiang2004preconditioned}, and earth surface reconstruction from seismic waveform tomography~\cite{rao2017seismic}. The local optimiser \emph{principal axis} (PRAXIS)~\cite{brent2013algorithms} is applied for the evaluation of image compression~\cite{950170}, finite element modelling in the biomedical ultrasound industry for the design of piezoelectric transducers~\cite{reynolds2004application}, and estimation of accuracy of dichotomous tests in the psychometric class~\cite{unlu2006estimation}. The \emph{improved stochastic ranking evolution strategy} (ISRES)~\cite{runarsson2000stochastic} is applied for adjusting high energy resolution X-ray beamlines~\cite{zhan2021development} and modelling parameters for determining fatigue crack growth in novel materials~\cite{iliopoulos2023framework}. The global optimiser DIRECT-L~\cite{gablonsky2000locally} is used for the modelling of metal-fill parasitic capacitance to reduce manufacturing defects of on-chip transmission lines~\cite{shilimkar2017modeling} and design of analogue integrated circuits~\cite{nicosia2008evolutionary}. Modern architecture uses high-performance computing (HPC) to solve complex minimisation problems on high-resolution data sets through access to large memory and high computational power. It becomes relevant to analyse the characteristics and performance of HPC hardware (e.g., heterogeneous clusters) and software (e.g., PETSc~\cite{balay2019petsc}, SLEPc~\cite{hernandez2005slepc}) for solving different minimisation problems in signal processing applications.

Given an input signal~$f: \Omega \rightarrow\mathbb{R}$ defined on a connected and compact domain~$\Omega$ in~$\mathbb{R}^d$, several problems in signal processing are associated with the computation of an approximating function as the solution to the minimisation problem
\begin{equation}\label{EQ:MINIMISATION}
\left\{\begin{matrix*}[l]
\min_{\mu} \| f-\hat{f}(\mu) \|_2^2 + \alpha \mathcal{P}(\hat{f}(\mu),\\
 \text{s.t.} \quad g(\mu),
\end{matrix*}\right.
\end{equation}
where the approximating function~$\hat{f}$ depends on a set of variables~$\mu$ to be optimised;~$\mathcal{P}$ is a penalisation operator that may be applied to regularise the approximating function;~$\alpha$ is a scalar coefficient; and \mbox{$g(\cdot)$} is a set of constraints. Starting from the general formulation in Eq. (\ref{EQ:MINIMISATION}), we analyse two minimisation problems in signal processing that require computationally expensive algebraic operations: (i) an approximation problem, where the functional is non-convex, non-linear, and the derivatives in the analytic form are not available; (ii) a denoising problem, where analytic derivatives are known, with bound constraints. Both these problems are computationally attractive: the availability or otherwise of the analytical derivatives and the presence of non-linear constraints affect the selection of the optimiser. Furthermore, the properties of each functional (e.g., convexity, algebraic operations) induce a large number of iterations of the minimisation method and, consequently, evaluations of the functional. In this context, the efficient computation of the functional drastically reduces the execution time of the minimisation problem. Exploiting the advantages of HPC hardware and software allows the user to improve the computational time of the minimisation.

As the main contribution, we discuss the properties of convergence, execution time, and scalability of five minimisation methods, i.e., the global optimisers DIRECT-L~\cite{gablonsky2000locally} and \emph{improved stochastic ranking evolution strategy} (ISRES)~\cite{runarsson2000stochastic}, and the local optimisers \emph{principal axis} (PRAXIS)~\cite{brent2013algorithms}, \emph{Limited-memory Broyden, Fletcher, Goldfarb, Shanno} (L-BFGS)~\cite{zhu1997algorithm}, and \emph{constrained optimisation by linear approximations} (COBYLA)~\cite{Powell1994}. We perform an efficient implementation of the proposed problems with HPC techniques, a comparison of the minimisation solvers to find the optimal solution, and a discussion of the main results in terms of execution time, scalability, and convergence for both the approximation~(Sect.~\ref{SEC:PROBLEM1}) and denoising~(Sect.~\ref{SEC:PROBLEM2}) problems.  These problems apply the main algebraic operations common to most minimisation problems in signal processing, and our analysis can be extended to other classes of problems. Finally, we show some possible results of the proposed problems, discussing conclusions and future work~(Sect.~\ref{SEC:CONCLUSIONS}).

\section{Related work\label{SEC:RELATEDWORK}}
We discuss the main methods for the solution of minimisation problems, their computational cost, available scientific libraries and hardware.

\paragraph{Minimisation solvers\label{SEC:OPTIMISATION}}
DIRECT~\cite{jones1993lipschitzian} is a global, derivative-free, and deterministic search algorithm that systematically divides the search domain into smaller hyperrectangles. Rescaling the bound constraints to a hypercube gives equal weight to all dimensions in the search procedure. DIRECT derives from Lipschitzian global optimisation, i.e., a branch-and-bound model where bounds are computed through the knowledge of a Lipschitz constant for the objective function. DIRECT introduces modifications to the Lipschitzian approach to improve the results in higher dimensions by eliminating the need to know the Lipschitz constant. The global optimisers DIRECT-L~\cite{gablonsky2000locally} is the locally-biased form that improves the efficiency of functions without too many local minima. As a global method, DIRECT-L spans all the possible solutions without finding local minima as the optimal solution. Furthermore, analytic or numeric derivatives are unnecessary to compute the optimal solution. Finally, DIRECT-L supports unconstrained and linearly-constrained problems.

The \emph{improved stochastic ranking evolution strategy} (ISRES)~\cite{runarsson2000stochastic} balances between objective and penalty functions stochastically, i.e., stochastic ranking, by proposing an improved evolutionary algorithm that accounts for evolution strategies and differential variation. The evolution strategy combines a mutation rule (with a log-normal step-size update and exponential smoothing) and differential variation (a Nelder–Mead-like update rule). The fitness ranking is simplified through the objective function for problems without non-linear constraints, while when non-linear constraints are included, the stochastic ranking is employed. ISRES supports unconstrained and constrained problems.

The local optimiser \emph{principal axis} (PRAXIS)~\cite{brent2013algorithms} is a gradient-free local optimiser that minimises a multivariate function through the \emph{principal-axis} method. PRAXIS is a modification of Powell's direction-set method~\cite{powell1964efficient}; given~$n$ variables, the set of search directions~$n$ is repeatedly updated until a set of conjugate directions with respect to a quadratic form is reached after~$n$ iterations. To ensure the correctness of the minimum, the matrix of the search directions is replaced by its principal axes so that the direction set spans the entire parameter space. PRAXIS is designed for unconstrained problems; bound constraints can be applied by considering a penalisation when constraints are violated.

The \emph{Limited-memory Broyden, Fletcher, Goldfarb, Shanno} (L-BFGS)~\cite{zhu1997algorithm} is an optimisation algorithm in the family of quasi-Newton methods that approximates the \emph{Broyden-Fletcher-Goldfarb-Shanno algorithm} (BFGS) using limited computer memory. Analogously to BFGS, the L-BFGS solver estimates the inverse Hessian matrix for the minimum search in the variable space; however, the L-BFGS method represents the approximation through a few vectors, thus involving a limited memory requirement. At each iteration, a short history of the past updates of the position and the gradient of the energy functional is used to identify the direction of the steepest descent and to implicitly perform operations requiring vector products with the inverse Hessian matrix. Since L-BFGS needs the derivatives of the functional, they are computed through numerical methods (e.g., finite difference method), where they are not available in analytic terms. L-BFGS supports both unconstrained and constrained problems.

The \emph{constrained optimisation by linear approximations} (COBYLA)~\cite{Powell1994} generates successive linear approximations of the objective function and constraints through a simplex of~$n+1$ points in~$n$ dimensions and optimises these approximations in a trust region at each step. Each iteration defines linear approximations of the objective and constraint functions by interpolating at the vertices of the simplex, and a trust region bound restricts each change to the variables. A new vector of variables is computed, which may replace one of the current vertices, either to improve the shape of the simplex or because it provides the best results according to a merit function that accounts for the constraint violation. For a deeper review of optimisation methods, we refer the reader to the survey in~\cite{rios2013derivative}.

\paragraph{Memory storage and computational cost}
Given a set of~$n$ variables, the L-BFGS method requires a memory storage of~$\mathcal{O}(n^{2})$ and the computational cost is~$\mathcal{O}(n v)$ at each iteration, where~$v$ is the number of steps stored in memory. For the PRAXIS method, the computational cost is~$\mathcal{O}(n^2)$. For the DIRECT-L method, the worst case computational cost is ~$\mathcal{O}(2^n)$, even if novel variants of this method propose improved performances. COBYLA has~$\mathcal{O}(n^2)$ computational cost. The population size for ISRES is~$20\cdot(n+1)$. Finally, we apply global and local optimisers consecutively: the global optimiser searches for the solution to the minimisation problem in the global parameter space. In contrast, the local optimiser refines the accuracy of the solution. 

\paragraph{Numerical libraries and hardware\label{SEC:HARDWARE}}
Among scientific libraries, we apply Eigen~\cite{eigenweb} for the data structures management, BLAS~\cite{blackford2002updated} and sparse BLAS~\cite{155391} for the dense and sparse matrices operations, PETSc~\cite{balay2019petsc} and SLEPc~\cite{hernandez2005slepc} for the parallel interface to BLAS routines, IBM Spectrum MPI~\cite{urlibm} for the interface to distributed computing environments. Tests and analyses are performed on CINECA cluster Marconi100. The CINECA Marconi100 cluster occupies the 26th position in the ``\emph{top500}'' list~\cite{urlcineca}. The cluster uses 980 nodes, each with IBM Power9 AC922 at 3.1GHz 32 cores and 4 NVIDIA Volta V100 GPUs per node, with the GPU interconnection NVlink 2.0 at 16GB and 256GB of RAM each node.

\section{Constrained least-squares approximation\label{SEC:PROBLEM1}}
The approximation of an input signal is widespread in image processing, e.g., computer graphics for compression~\cite{dhawan2011review} and restoration~\cite{banham1997digital}, biomedicine~\cite{mccann2019biomedical} and physics~\cite{thiebaut2009image} applications, and graph processing, e.g., time-varying graphs signal reconstruction~\cite{qiu2017time} and compression~\cite{besta2019slim}.
Given the problem in Eq.~(\ref{EQ:MINIMISATION}), we analyse the minimisation problem for the signal approximation (Sect.~\ref{SEC:MODELRECONSTRUCTION}) and discuss the experimental results (Sect.~\ref{SEC:UNC}).
\subsection{Constrained least-squares approximation\label{SEC:MODELRECONSTRUCTION}}
Given an input signal \mbox{$f:\Omega\rightarrow\mathbb{R}$} and a set \mbox{$\mathcal{B} = \{\varphi_j \}_{j=1}^{n}$} of basis functions, we consider the approximating function \mbox{$\hat{f}:=\sum_{j=1}^{n}\beta_{j}\varphi_{j}$} with \mbox{$\beta:=(\beta_{j})_{j=1}^{n}$}.
In our setting, we assume that \mbox{$\varphi_{j}(\mathbf{q}):=\varphi(\|\mathbf{q}-\mu_{j}\|_{2})$} is a radial basis function (RBF) generated by a 1D function \mbox{$\varphi:\mathbb{R}\rightarrow\mathbb{R}$} (e.g., \mbox{$\varphi(s):=\exp(-s)$}) and~$\mu_{j}$ is its centre. To compute the approximating function, we solve the constrained minimisation problem in Eq. (\ref{EQ:MINIMISATION}), where the variables are the centres \mbox{$\mu:=(\mu_{j})_{j=1}^{n}$} of the RBFs, \mbox{$\alpha:=0$}, and the constraints \mbox{$g(\mu)$} represent the membership of the centres~$\mu_{j}$ to~$\Omega$, as
\begin{equation}
\label{EQ:reconstructionRegular}
\left\{\begin{matrix*}[l]
\min_{\mathbf{\mu}} \| f- \hat{f}(\mu) \|^2,
\\
\text{s.t.} \ \mu_{j}\in\Omega,\quad j=1,\ldots,n.
\end{matrix*}\right.
\end{equation}
In the discrete setting, we sample~$f$ and~$\hat{f}$ at a set \mbox{$\mathcal{Q}:=\{\mathbf{q}_{i}\}_{i=1}^{m}\subseteq \Omega$} of points, i.e.,~$\mathbf{f}:=(f(\mathbf{q}_{i}))_{i=1}^{m}$ and~$\hat{\mathbf{f}}(\mu):=(\hat{f}(\mathbf{q}_{i}))_{i=1}^{m}$. The approximating signal is defined as
\begin{equation}
\label{eq:APPROX-PROBLEM}
\hat{f}(\mathbf{q}_{i}):=\sum_{j=1}^{n}\beta_{j}\varphi(\|\mathbf{q}_{i}-\mu_{j}\|_{2}),\quad i=1,\ldots,m,
\end{equation}
i.e.,~$\hat{\mathbf{f}}(\mu)=\mathbf{\Phi}(\mu)\beta(\mu)$, where \mbox{$\mathbf{\Phi}(\mu):=(\varphi(\mathbf{q}_i,\mu_j))_{i=1 \dots m}^{j=1 \dots n}$} is the \mbox{$m\times n$} Gram matrix. Since \mbox{$n<<m$}, the solution to \mbox{$\mathbf{\Phi}^{\top}(\mu)\mathbf{\Phi}(\mu)\beta(\mu) = \mathbf{\Phi}^\top(\mu)\mathbf{f}$} in Eq. (\ref{EQ:reconstructionRegular}) is generally ill-conditioned; indeed, we solve the regularised linear system
\begin{equation}
\label{EQ:LINSYS}
(\mathbf{\Phi}^\top(\mu) \mathbf{\Phi}(\mu) + \lambda \mathbf{I})\beta(\mu)=\mathbf{\Phi}^\top(\mu)\mathbf{f},
\end{equation}
with~$\lambda = 1e-12$ and constraints \mbox{$\mu_{j}\in\Omega$,~$j=1,\ldots,n$}.

We discuss two variants of this problem: the first one is the bound-constraint version, which is typical of image processing as the signal is defined on a regular grid~\cite{cammarasana2021kernel}; the second one is the non-linear geometric constraint version, which is typical of signals on graphs/meshes~\cite{paccini2022three}. The two variants share the same objective function; however, the non-linear geometric constraints affect the selection and analysis of the minimisation method.
\begin{algorithm}[t]
\caption{Constrained least-squares approximation.\label{ALG:FUNCTIONAL}}
\begin{algorithmic}[1]
\State~$\mathbf{f}=$ Input discrete signal
\Procedure{$\hat{\mathbf{f}}$ = approximate}{$\mathbf{f}$}
\State Mat~$\mathbf{\Phi}(\mu)$
\State Mat~$\mathbf{\Phi}_T(\mu) = \mathbf{\Phi}^\top(\mu)$
\State Mat~$\overline{\mathbf{\Phi}}(\mu)=\mathbf{\Phi}_T(\mu) \mathbf{\Phi}(\mu)$
\State Mat~$\mathbf{A}(\mu)=\overline{\mathbf{\Phi}}(\mu) +\lambda \mathbf{I}$
\State Vec~$\mathbf{b}(\mu)=\mathbf{\Phi}(\mu) \mathbf{f}$
\State Vec~$\beta(\mu) = \mathbf{A}(\mu) \setminus \mathbf{b}(\mu)~$
\State Vec~$\hat{\mathbf{f}}(\mu):=\mathbf{\Phi}(\mu)\beta(\mu)$
\State Real~$\epsilon = \| \mathbf{f} - \hat{\mathbf{f}}(\mu) \|_2$
\EndProcedure
\State Apply constraints~$g(\mu)$:~$\mu_{j}\in\Omega$,~$j=1,\ldots,n$.
\end{algorithmic}
\end{algorithm}
\paragraph{Algorithm and parallelisation}
The objective function in Eq.~(\ref{EQ:reconstructionRegular}) is computed through the Algorithm~\ref{ALG:FUNCTIONAL}.
\begin{itemize}
\item Line 1 is performed out of the computation of the functional. One MPI process reads the input signal, scatters the signal values across the MPI processes, and broadcasts the input points~$m$.
\item Line 3 (\emph{$k-$nn search} and \emph{matrix definition}) computes the matrix~$\mathbf{\Phi}$. The sparsity of the matrix depends on the parameters of the~$k-$d search (e.g.,~$k$ index). We parallelise the query on the~$k-$d tree, where each MPI process is assigned with a sub-set of centres.
\item Line 4 (\emph{Matrix transpose}) computes the transpose of the matrix~$\mathbf{\Phi}$ that computationally corresponds to a sparse matrix copy (BLAS \emph{omatcopy}).
\item Line 5 (\emph{Matrix-matrix multiplication}) computes the sparse matrix-sparse matrix multiplication through a BLAS \emph{usmm} routine, with~$\mathcal{O}(k\cdot k \cdot m)$ operations, where~$k$ is the number of non-zero elements per row of the sparse matrix, and~$m$ is the input point set.
\item Line 6 (\emph{Matrix shift}) computes a matrix shift, which computationally corresponds to a BLAS \emph{axpy}, linear cost with~$m$.
\item Line 7 (\emph{Matrix-vector multiplication}) computes a sparse matrix-vector multiplication through a BLAS \emph{usmv} routine, with~$\mathcal{O}(2km)$ operations.
\item Line 8 (\emph{Solve system}) solves the sparse linear system. The computational cost depends on the selected algorithm.
\item Line 9 (\emph{Matrix-vector multiplication}) computes a sparse matrix-vector multiplication through a BLAS \emph{usmv} routine, with~$\mathcal{O}(2km)$ operations.
\item Line 10 (\emph{Vex AXPY} and \emph{Vec Norm}) computes the error norm through a BLAS vector sum (\emph{axpy}) and a BLAS vector norm (\emph{nrm2}), both with a computational cost that is linear with the number of input points.
\item Line 11 computes the non-linear constraints for the variables~$\mu$.
\end{itemize}
All the BLAS operations are parallelised by distributing the matrices and vectors by rows among the MPI processes. 
\begin{figure}[t]
\centering
\includegraphics[width=.7\columnwidth]{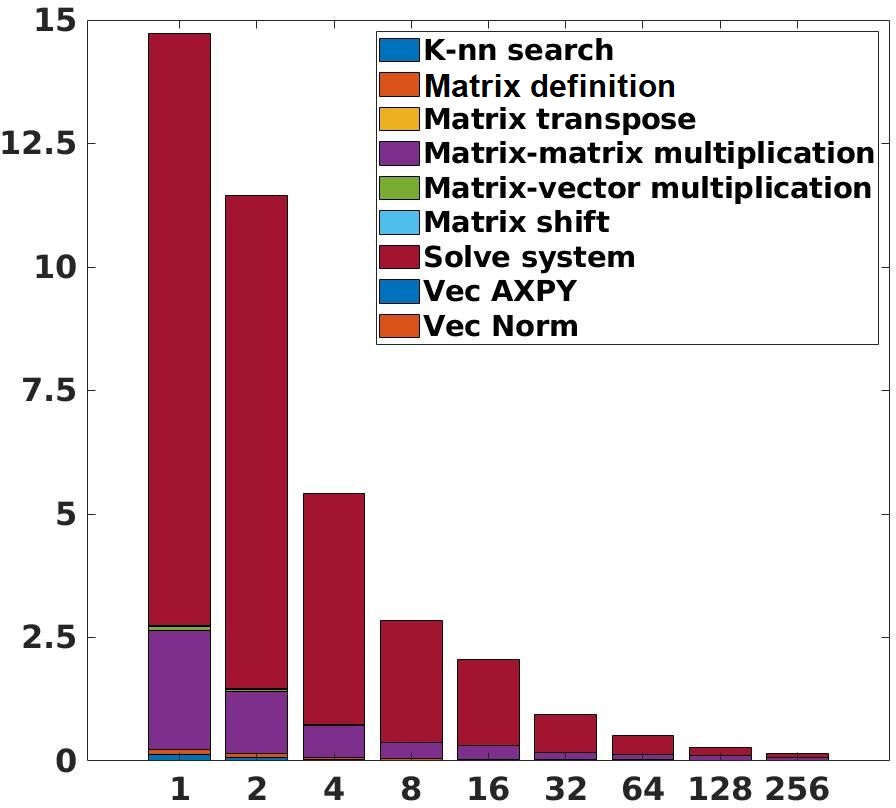}
\caption{Execution time ($y-$axis, in seconds) with respect to the processes ($x-$axis), approximation problem.\label{FIG:SCALABILITY}}
\end{figure}
\paragraph{Solution of the linear system}
The solution of the sparse linear system is the main operation of our problem in computational terms. According to~\cite{galizia2022evaluating}, we select as solver the iterative biconjugate gradient stabilised (BICGSTAB)~\cite{van1992bi} method with the Block-Jacobi preconditioner. The BICGSTAB is a transpose-free version of the biconjugate gradient (BICG) method~\cite{fletcher2006conjugate} that improves convergence and smoothness. BICGSTAB is composed of two matrix-vector multiplications, two scalar products, one vector norm, and two vector sum operations (i.e., \emph{waxpy} and \emph{axpbypcz}) that are linear with the elements of the vectors. The computational cost is~$\mathcal{O}(k m t)$ with~$t$ iterations,~$k$ non-zero elements of the sparse matrix, and~$m$ rows. BICGSTAB guarantees the stability of the solution and good scalability properties. 
\begin{figure*}[t]
\centering
\begin{tabular}{ccc}
\includegraphics[width=.4\textwidth]{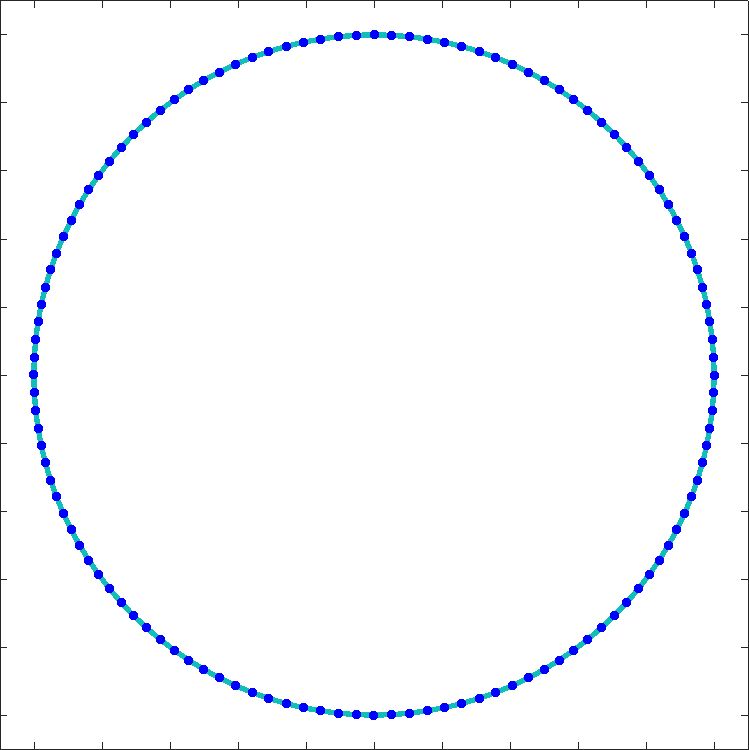} &
\includegraphics[width=.4\textwidth]{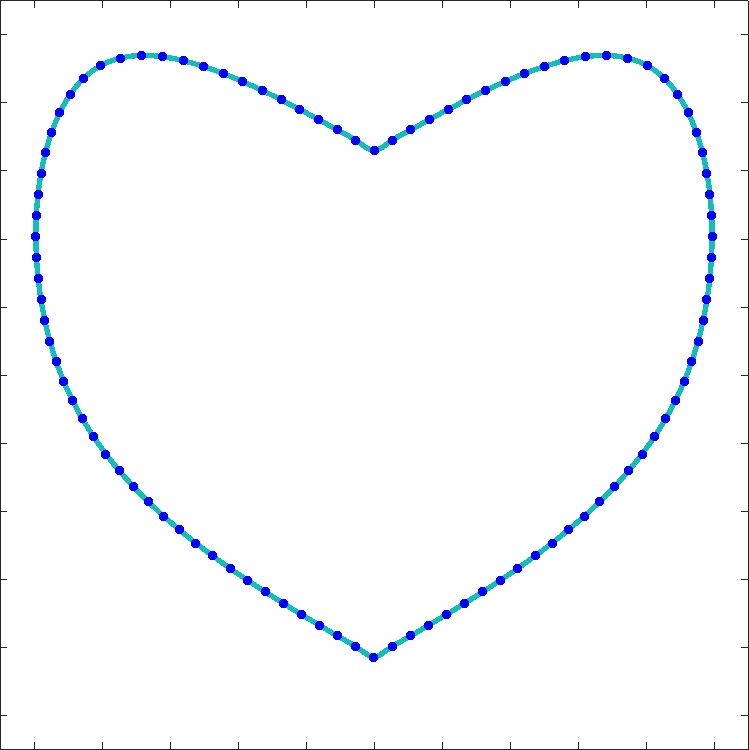} \\
\end{tabular}
\caption{Examples of input points sampled on a curve (blue dots) and the respective interpolating curve (cyan). \label{FIG:CURVES}}
\end{figure*}
\subsection{Experimental results\label{SEC:UNC}}
\paragraph{Bound-constraints experimental set-up}
We define an input signal discretised on a 2D regular grid of~$m\times m=65K$ points, with~$n=12K$ variables which are related to the 2D spatial coordinates of~$6K$ centres of the RBFs of the approximating function~$\hat{\mathbf{f}}$ in Eq. (\ref{eq:APPROX-PROBLEM}). The matrix~$\mathbf{\Phi}$  is computed with a~$k-$nearest neighbourhood of 200 points. The coefficient matrix of the linear system in Eq.~(\ref{EQ:LINSYS}) is a sparse matrix of~$65K \times 65K$ with~$40K$ non-zero elements.
The bound constraints force each variable to belong to the image domain, i.e., to fall between the lower and upper bounds for each of the two dimensions. 
\begin{table*}[t]
\centering
\caption{Comparison among minimisation methods of the approximation problem on 2D images. $t=1200s$ with 256 processes. The best results are in bold.\label{TAB:COMPARISON1}}
\begin{tabularx}{\textwidth}{p{0.34\textwidth}|p{0.22\textwidth}p{0.22\textwidth}p{0.22\textwidth}}
Metric& Functional value & Functional count & Time [s] \\ \hline
PRAXIS&$\mathbf{4.96}$ & 8K  &$t$ \\
DIRECT-L&26.5 & 2M & 40$t$ \\
L-BFGS& 13.94 & 0.5K  & 2$t$ \\ 
COBYLA& 9.48 & 8K  & 4$t$ \\
ISRES&$>30$ &$\mathbf{195}$  &$<t$ \\
DIRECT-L + PRAXIS& 25.94 & 2M + 7K  & 41$t$ \\
DIRECT-L + L-BFGS& 26.46 & 2M + 3K  & 42$t$
\end{tabularx}
\end{table*}
\paragraph{Comparison between minimisation methods}
Table~\ref{TAB:COMPARISON1} shows the comparison among minimisation methods for the solution of Eq.~(\ref{EQ:reconstructionRegular}) with bound-constraints. PRAXIS has the best performance both in terms of computation time and functional value. DIRECT-L has a very high computation time due to the global search for the optimal solution. L-BFGS does not converge to the optimal solution. Furthermore, the initialisation of the solution through DIRECT-L does not improve the minimisation of PRAXIS and L-BFGS. Finally, both COBYLA and ISRES have worse performance than PRAXIS: ISRES does not converge to an optimal solution, while COBYLA has worse results both in terms of functional value and execution time.

\paragraph{Scalability of functional computation}
Fig.~\ref{FIG:SCALABILITY} and Table~\ref{TAB:SCALABILITY} show the scalability of the Algorithm~\ref{ALG:FUNCTIONAL} without constraints. We mention that~$k-$nn search is a perfectly parallel operation. Matrix definition strictly depends on the type of the generating functions,~$k$-index value, and other parameters related to the type of application. Matrix-matrix multiplication and linear system solving are the most expensive operations and show good scalability when increasing the number of processes. In particular, the matrix-matrix multiplication passes from 2.4 seconds with 1 process to 0.06 seconds with 256 processes; the linear system solve passes from 12 seconds with 1 process to 0.07 seconds with 256 processes. The total time varies from 14.7 seconds with 1 process to 0.95 with 32 processes and 0.15 seconds with 256 processes. The efficiency is~$48\%$ with 32 processes and~$38\%$ with 256 processes. We recall that each node of the Marconi100 cluster is composed of 32 processes, and the efficiency further reduces when inter-node communications are required.
\begin{table*}[t]
\centering
\caption{Scalability analysis of each operation (Op.) in milliseconds of the approximation problem. \label{TAB:SCALABILITY}}
\begin{tabularx}{\textwidth}{p{0.02\textwidth}|p{0.0575\textwidth}p{0.0575\textwidth}p{0.0575\textwidth}p{0.0575\textwidth}p{0.0575\textwidth}p{0.0575\textwidth}p{0.0575\textwidth}p{0.07\textwidth}p{0.07\textwidth}p{0.0575\textwidth}}
Op. &K-nn search &Matrix def. & Mat transp. & Mat-Mat mult. & Mat-Vec mult. & Matrix shift & Solve system & Vec-Vec add. & Vec norm & Total \\ \hline
1 & 122 & 111 & 5 & 2407 & 73 & 27 & 11978 &~$<0.1$ &~$<0.1$ &14725 \\
2& 65 & 81 & 5 &1266 &39 &17 & 9966 &~$<0.1$ &~$<0.1$ & 11444\\
4& 32 & 35  & 5 & 639 & 23 & 8 & 4681 & ~$<0.1$ &~$<0.1$ & 5425\\
8& 16 & 29 & 5  & 321 & 8 & 4 & 2463 &~$<0.1$ &~$<0.1$ & 2850\\
16& 8 &19 & 5 & 275 & 4 & 5 & 1748 &~$<0.1$ &~$<0.1$ & 2070\\
32& 4 & 19 & 9 &132 & 4 & 2 &766 &~$<0.1$ &~$<0.1$ & 952\\
64&2&10&11&105&2&1&376&$<0.1$&$<0.1$&518\\
128&1&6&13&88&1&$0.4$&171&$<0.1$&$<0.1$&290\\
256& 0.2 & 2 & 3 & 60 & 0.8 & 0.7 & 71 &~$<0.1$ &~$<0.1$ &150\\
\end{tabularx}
\end{table*}
\begin{table*}[t]
\centering
\caption{Comparison among minimisation methods for the approximation problem with constraints. The best results are in bold. \label{TAB:COMPARISON3}}
\begin{tabularx}{\textwidth}{L|LLL}
Metric& Functional value & Functional count  & Time [s]  \\ \hline
ISRES&$\mathbf{4.05}$ &500K  &~$>3K$\\
COBYLA& 4.51 & 44K &120\\
L-BFGS&4.84 &$\mathbf{1500}$  & 5
\end{tabularx}
\end{table*}
\paragraph{Non-linear geometric constraints}
To consider curved domains (e.g., Fig.~\ref{FIG:CURVES}), we include a non-linear constraint that accounts for the geometry of the domain for each variable~$\mu_i$ of the minimisation problem in Eq.~(\ref{EQ:reconstructionRegular}). In particular, given the set of~$m$ input points representing a discrete curve, we compute the interpolating curve and define the non-linear constraints as an equality constraint of the distance between each variable and the curve. Fig.~\ref{FIG:CURVES} shows examples of input points sampled on a curve and the respective interpolating curve. In our tests, we select a signal defined on a curve discretised with 2K input points and optimise 125 variables~$\mu_i$.
\begin{figure}[t]
\centering
\includegraphics[width=.69\columnwidth]{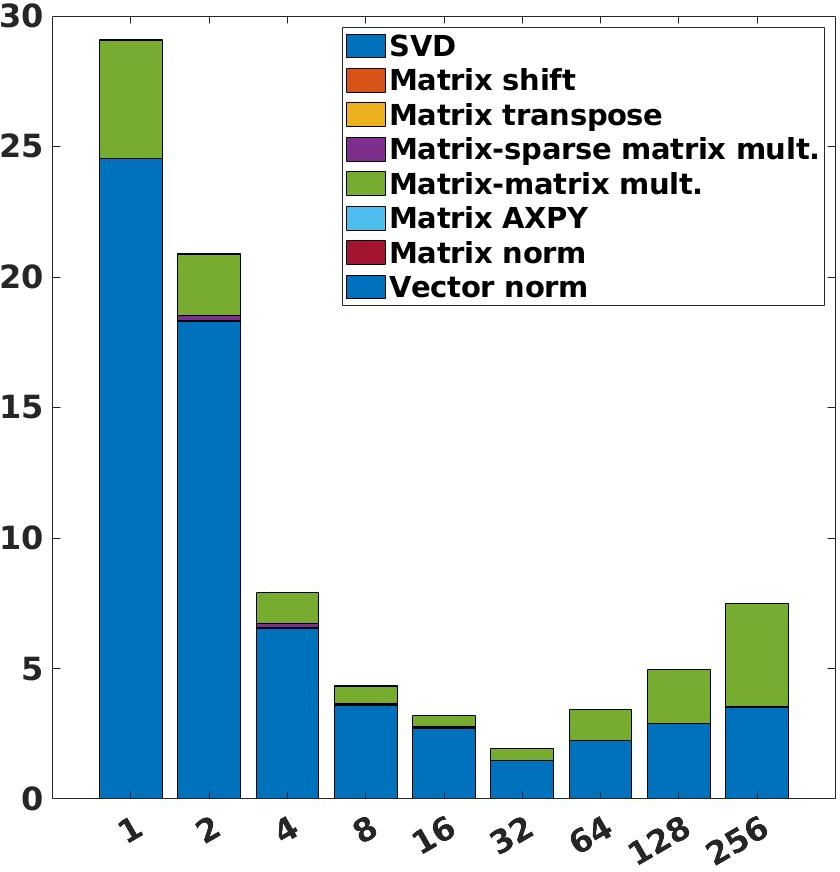} 
\caption{Execution time ($y-$axis, in seconds) with respect to the processes ($x-$axis), denoising problem.\label{FIG:SCALABILITY2}}
\end{figure}
\paragraph{Comparison among minimisation methods}
For the computation and scalability of the functional, we refer to the Algorithm~\ref{ALG:FUNCTIONAL}. In this case, the non-linear constraint affects the selection of the minimisation solver. In Table~\ref{TAB:COMPARISON3}, we discuss the convergence of the minimisation methods for the solution of Eq.~(\ref{EQ:reconstructionRegular}): L-BFGS does not converge to the optimal solution, and the global optimiser ISRES has better accuracy with respect to the local optimiser COBYLA (4.05 vs 4.51), at the cost of a larger number of iterations and execution time. Finally, PRAXIS and DIRECT-L do not manage non-linear constraints and can not be applied.
\begin{algorithm}[t]
\caption{Constrained SVD for signal denoising.\label{ALG:FUNCTIONAL2}}
\begin{algorithmic}[1]
\State~$\mathbf{f}=$ Input discrete signal
\Procedure{$\hat{\mathbf{f}}$ = denoise}{$\mathbf{f}$}
\State Mat~$ \mathbf{U}\mathbf{S}\mathbf{V}$ = SVD($\mathbf{f}$)
\State Mat~$\hat{\mathbf{S}} = \mathbf{S} - \mu$
\State Mat~$\overline{\mathbf{U}} = \mathbf{U} \hat{\mathbf{S}}$
\State Mat~$\mathbf{V}_T =  \mathbf{V}^\top$
\State Mat~$\hat{\mathbf{f}} = \overline{\mathbf{U}}\mathbf{V}_T$
\State Real~$\epsilon_1 = \| \mathbf{f} -  \hat{\mathbf{f}} \|_F~$
\State Real~$\epsilon_2 = \| \hat{\mathbf{S}}  \|_1~$
\State Real~$\epsilon =\epsilon_1 + \alpha \epsilon_2$
\EndProcedure\\
Apply constraints $g(\mu)$:~$\mathbf{S}_{ii} - \mu_i > 0$,~$i=1,\ldots,m$
\end{algorithmic}
\end{algorithm}
\section{Constrained SVD for signal denoising\label{SEC:PROBLEM2}}
The denoising of signals is widespread in many applications; images acquired by digital sensors are generally affected by different types of noise, such as speckle~\cite{burckhardt1978speckle} and exponential~\cite{shrivastava2007approach} noise on biomedical images, salt-and-pepper~\cite{azzeh2018salt}, Gaussian~\cite{russo2003method}, and Poisson~\cite{trussell2012dominance} noise on images acquired through camera sensors. Given the problem in Eq.~(\ref{EQ:MINIMISATION}), we analyse the minimisation for the denoising problem (Sect.~\ref{SEC:MODELDENOISE}) and discuss the experimental results (Sect.~\ref{SEC:TESTDENOISE}).

\subsection{Constrained SVD\label{SEC:MODELDENOISE}}
Given an input 2D image~$\mathbf{f}$ represented on a squared~\mbox{$m \times m$} grid, we compute the singular values decomposition~\mbox{$\mathbf{f} =  \mathbf{U}\mathbf{S}\mathbf{V}^\top$} where~$\mathbf{U}$ and~$\mathbf{V}$ are~$m \times m$ dense matrices and~$\mathbf{S}$ is a diagonal~$m \times m$ matrix. This factorisation is well-known for separating high-frequency by low-frequency components of the image and allowing us to reduce the noise components while preserving the features/properties of the input image. We define the approximating function~$\hat{\mathbf{f}} = \mathbf{U}\hat{\mathbf{S}}\mathbf{V}^\top$, where~$\hat{\mathbf{S}}$ is computed through a threshold operation on~$\mathbf{S}$. The penalisation term is defined as the nuclear norm of the approximated signal~$\mathcal{P} = \| \hat{\mathbf{f}} \|_{\ast}$; the nuclear norm is linked with the~$\mathbf{S}$ matrix of the SVD and regularises high-frequency components~\cite{cammarasana2023learning}.

We define the variables of our minimisation problem~$\mu = (\mu_i)_{i=1}^{m}$ as the threshold values to be applied to the diagonal of~$\mathbf{S}$ and we compute the singular values with applied the threshold values as~$\hat{\mathbf{S}} = \mathbf{S} - \mu$, where we assume \mbox{$\mathbf{S}-\mu$ as~$\mathbf{S}_{ii} -\mu_i$}, with~$\mathbf{S}_{ii}$ as the~$(i, i)$ entry of~$\mathbf{S}$. We add the bound constraint to the minimisation problem to ensure the non-negativity of the threshold singular values. We define the minimisation problem as
\begin{equation}
\label{EQ:denoisingRegular}
\left\{\begin{matrix*}[l]
\min_{\mu} \| \mathbf{f}-\mathbf{U}(\mathbf{S}-\mu)\mathbf{V}^\top \|_{F}^2 + \alpha \sum{(\mathbf{S} - \mu)},
 \\
 \text{s.t.} \quad \mathbf{S} - \mu > \mathbf{0}.
\end{matrix*}\right.
\end{equation}
This problem can be assumed as convex, and the derivatives in the analytic form are available. The computation of the objective function requires the application of the main algebraic operations, including BLAS~1 (e.g., \emph{axpy}), BLAS~3 (e.g., \emph{gemm}) and sparse BLAS~3 (e.g., \emph{usmm}). Furthermore, it includes the computation of the singular values of a dense/sparse matrix, depending on the selected input signal. The analysis of this problem is general and can be extended to other image-processing applications.
\begin{figure*}[t]
\centering
\begin{tabular}{ccc}
\includegraphics[width=.3\textwidth]{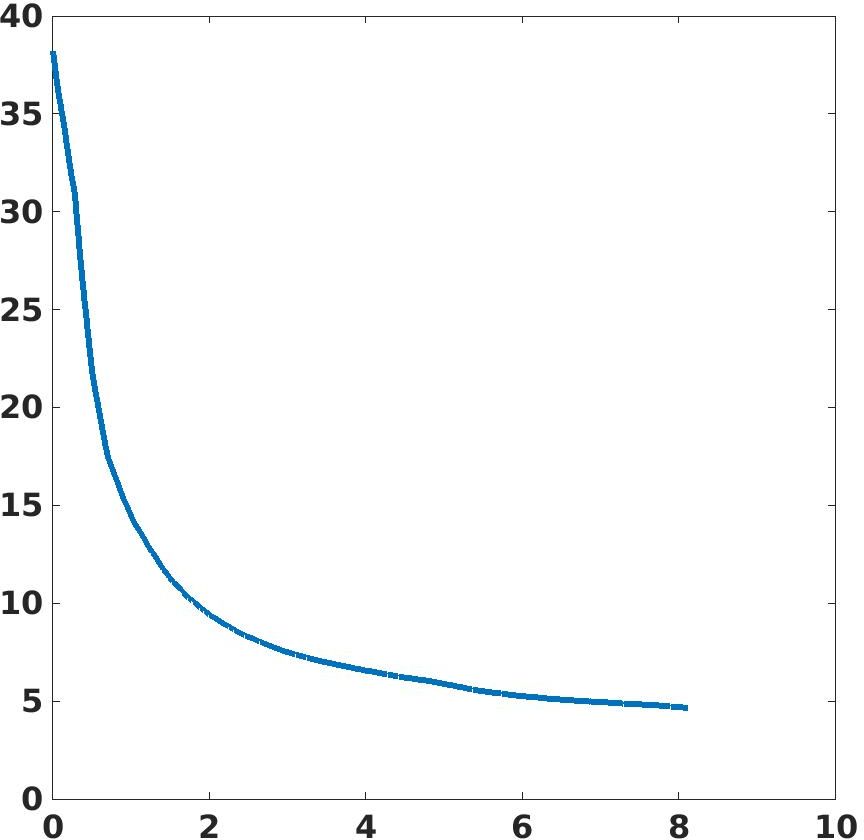} &
\includegraphics[width=.3\textwidth]{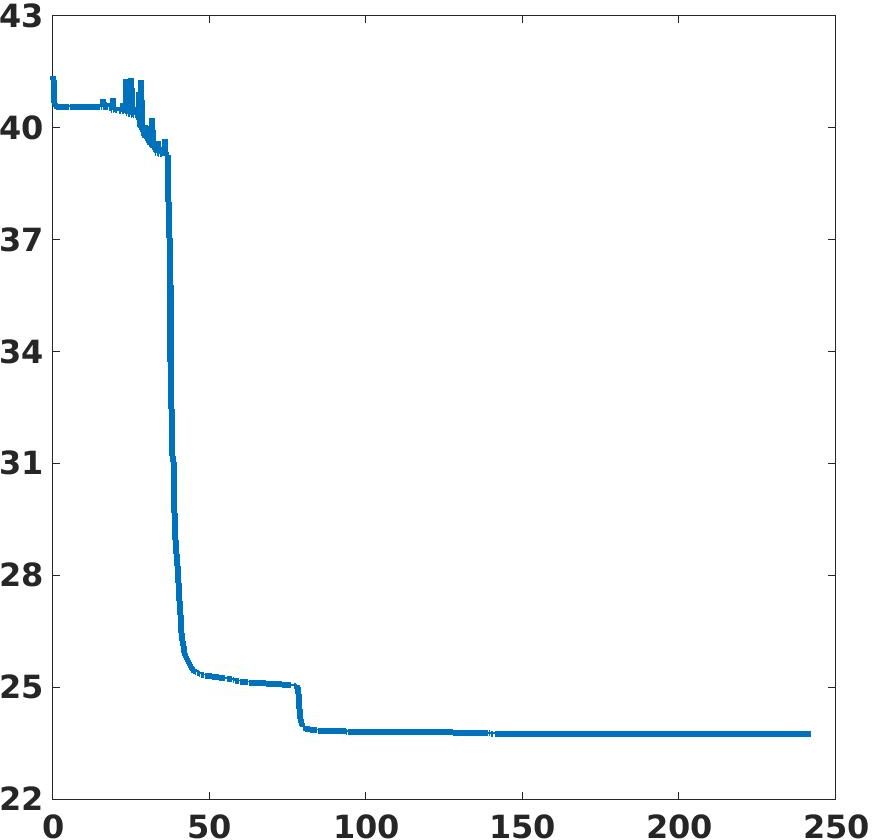} &
\includegraphics[width=.3\textwidth]{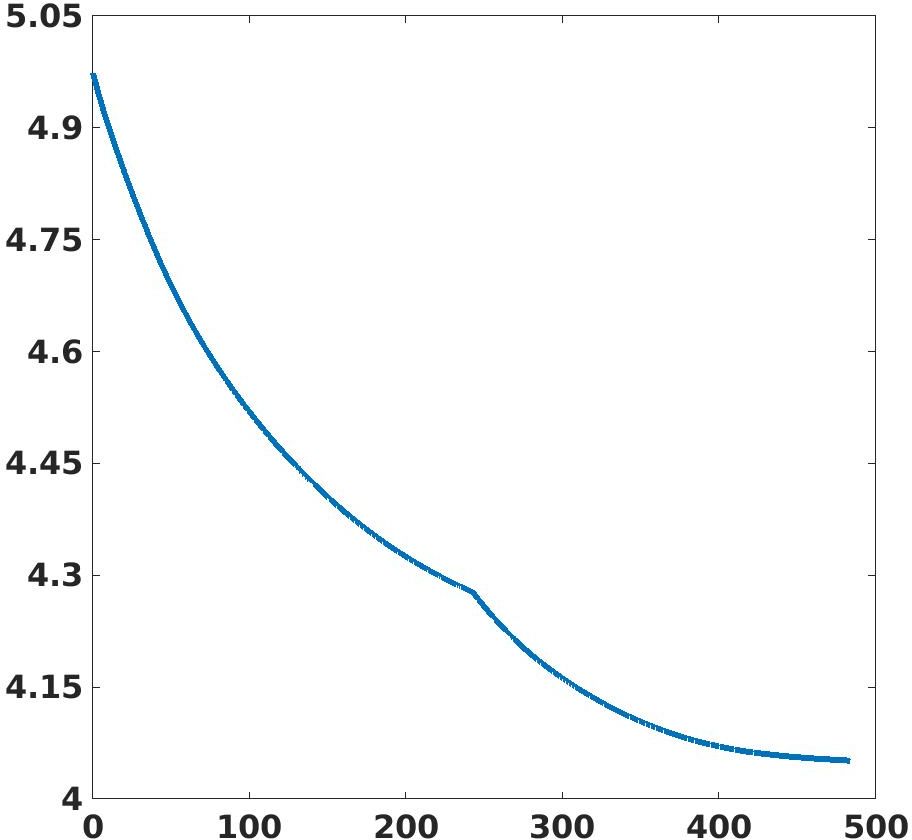} \\
(a) & (b) & (c) 
\end{tabular}
\caption{Minimisation of PRAXIS with functional value ($y-$axis) with respect to the number of evaluations of the functional ($x-$axis,~$\times 10^3$): (a) approximation problem, (b) denoising problem. Minimisation of ISRES, approximation problem with non-linear constraints (c). \label{FIG:CONVERGENCE}}
\end{figure*}
\begin{table*}[t]
\centering
\caption{Comparison among minimisation methods for the denoising problem on the 2D image ($5616 \times 3744$ matrix). $t=800s$ with 32 processes. The best results are in bold. \label{TAB:COMPARISON2METHOD}}
\begin{tabularx}{\textwidth}{p{0.34\textwidth}|p{0.22\textwidth}p{0.22\textwidth}p{0.22\textwidth}}
Metric &  Functional value & Functional count & Time [s]  \\ \hline
L-BFGS& 31.25  &$\mathbf{3}$ &$t$ \\ 
DIRECT-L&25.45 & 1M  &  2200$t$\\
PRAXIS&$\mathbf{24.47}$ & 250K  & 600$t$  \\
COBYLA&39.88 & 250K &1100$t$ \\
ISRES&$>50$ & 1M  & 2000$t$ \\
DIRECT-L + PRAXIS& 25.42 & 1M + 1K   &2210$t$ \\
DIRECT-L + L-BFGS& 25.45 & 1M  & 2200$t$
\end{tabularx}
\end{table*}
For the efficient computation of the SVD, the tridiagonalisation of the cross product matrix without forming it explicitly is achieved through the bidiagonalisation~$\mathbf{f} = \mathbf{P}\mathbf{B}\mathbf{Q}^*$ where~$\mathbf{P}$ and~$\mathbf{Q}$ are unitary matrices and~$\mathbf{B}$ is an upper bidiagonal matrix. Then, the SVD of~$\mathbf{B}$ is applied to recover the SVD of~$\mathbf{f}$. The bidiagonalisation is achieved through the Lanczos method~\cite{golub1965calculating}. According to our experimental tests (Sect.~\ref{SEC:TESTDENOISE}), we select the tridiagonalisation of the cross-product matrix as the SVD method.
\paragraph{Algorithm and parallelisation}
The objective function in Eq.~(\ref{EQ:denoisingRegular}) is computed through the Algorithm~\ref{ALG:FUNCTIONAL2}.
\begin{itemize}
\item Line 1 is performed out of the computation of the functional. One MPI process reads the input 2D signal, scatters the signal values across the MPI processes, and broadcasts the input points~$m$.
\item Line 3 (\emph{SVD}) computes the SVD decomposition of the image and saves two full matrices ($\mathbf{U}$,~$\mathbf{V}$) and a sparse matrix ($\mathbf{S}$).
\item Line 4 (\emph{Matrix shift}) computes a matrix shift, which computationally corresponds to a BLAS \emph{axpy}, linear cost with~$m$.
\item Line 5 (\emph{Matrix-sparse matrix multiplication}) computes a sparse matrix-matrix multiplication through sparse BLAS \emph{usmm} routine with~$\mathcal{O}(k\cdot m^2)$ operations, where~$k$ is the number of non-zero elements per row of the sparse matrix, and~$m$ is the input point set.
\item Line 6 (\emph{Matrix transpose}) computes the transpose of the right eigenvectors matrix that computationally corresponds to a matrix copy (BLAS \emph{omatcopy}).
\item Line 7 (\emph{Matrix-matrix multiplication}) computes a matrix-matrix product through BLAS \emph{gemm}, with a maximum computational cost of~$\mathcal{O}(m^3)$.
\item Line 8 (\emph{Matrix AXPY} and \emph{Matrix norm}) computes both a matrix-matrix addition and a matrix Frobenius norm in linear cost with the number of input points.
\item Line 9 (\emph{Vector norm}) computes the~$\mathcal{L}^1$ norm of the sparse matrix~$\hat{\mathbf{S}}$, linear cost with the number of variables.
\item Line 10 computes the sum of two scalars.
\item Line 12 computes the set of non-linear constraints for the variables~$\mu$, as~$\mu_i < \mathbf{S}_{ii}$ for each variable~$i$ with respect to the related diagonal entry of the singular values matrix~$\mathbf{S}$.
\end{itemize}
All the BLAS operations are parallelised by distributing the matrices and vectors by rows among the MPI processes.
\begin{table*}[t]
\centering
\caption{Comparison among SVD methods on dense~$5616 \times 3744$ matrix with first 300 singular values and sparse~$16368 \times 16384$ matrix. Execution time is expressed in milliseconds. N.C. means the method does not converge. \label{TAB:COMPARISON2}}
\begin{tabularx}{\textwidth}{p{0.13\textwidth}|p{0.116\textwidth}p{0.116\textwidth}p{0.116\textwidth} | p{0.116\textwidth}p{0.116\textwidth}p{0.116\textwidth}}
Matrix &\multicolumn{3}{c}{Dense} & \multicolumn{3}{c}{Sparse}  \\ 
Processes & 1 & 32 & 128 & 1 & 32 & 128 \\ \hline
Cross & 24548  & 1460 & 2880 & 19360 & 876 & 242 \\  
Randomized &\multicolumn{3}{c}{N.C.} & 90950 & 12543 & 9124  \\
Cyclic &~$>100K$ &~$>100K$ & 4217 &~$>100K$ & 12362 & 6855 \\
Lanczos &  \multicolumn{3}{c}{N.C.} & 28702 & 1160 & 306    \\
Trlanczos &6073  & 3415 & 3879 & 28599 &  1175& 302  \\
\end{tabularx}
\end{table*}
\subsection{Experimental results: bound constraints\label{SEC:TESTDENOISE}}
\paragraph{Experimental set-up}
We select an input signal as a high-resolution image or a weighted Laplacian matrix of a large grid. As a dense rectangular matrix, a 21-megapixel camera sensor acquires a typical image resolution of~$5616 \times 3744$. Given a regular grid of~$256 \times 256$, the Laplacian matrix is a~$16368 \times 16384$ matrix and is typically banded sparse. 
\begin{table*}[t]
\centering
\caption{Scalability analysis of each operation (Op.) in milliseconds of the denoising problem. \label{TAB:SCALABILITY2}}
\begin{tabularx}{\textwidth}{p{0.02\textwidth}|p{0.07\textwidth}p{0.07\textwidth}p{0.065\textwidth}p{0.065\textwidth}p{0.065\textwidth}p{0.065\textwidth}p{0.065\textwidth}p{0.07\textwidth}p{0.07\textwidth}}
Op. &SVD & Mat shift &  Mat transp. & Mat-Mat mult  & Mat-Mat mult & Mat-Mat add. & Mat norm & Vec norm  & Total \\ \hline
1 & 24548  &~$<0.1$ & 1& 2& 4503 & 21& 33&~$<0.1$ & 29111  \\
2 & 18293 &~$<0.1$ & 56 & 168 & 2340 & 13& 17&$<0.1$&20888 \\
4 & 6546 &$<0.1$&40&139&1168&6&8&$<0.1$&7910 \\
8 & 3580 &$<0.1$&23&73&648&4&4&$<0.1$&4334 \\ 
16 & 2716 &$<0.1$&14&41&419&3&2&$<0.1$&3197 \\
32 & 1460&$<0.1$&7&27&449&1&1&$<0.1$&1948\\
64 & 2223&$<0.1$&4&18&1189&$<0.1$&1&$<0.1$& 3436\\
128 & 2880&$<0.1$&4&11&2069&$<0.1$&$<0.1$&$<0.1$&4965\\
256 & 3517&$<0.1$&4&11&3945&$<0.1$&$<0.1$&$<0.1$& 7487
\end{tabularx}
\end{table*}
\paragraph{Comparison between minimisation methods}
Table~\ref{TAB:COMPARISON2METHOD} compares the minimisation methods for the solution of Eq.~(\ref{EQ:denoisingRegular}). L-BFGS converges to the global minimum and has the best results in terms of execution time/functional evaluation number; the knowledge of the analytic derivatives allows the method to compute the optimal minima with few evaluations of the functional. PRAXIS and DIRECT-L have a large number of evaluations of the functional. However, they both perform better than L-BFGS even without knowing the derivatives. In particular, PRAXIS has better results than DIRECT-L. Finally, the initialisation of the solution through DIRECT-L does not improve the minimisation of PRAXIS and L-BFGS. COBYLA and ISRES have worse results than PRAXIS regarding the functional value and computation time.
\begin{figure*}[t]
\centering
\begin{tabular}{ccc}
\includegraphics[width=.3\textwidth]{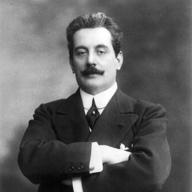} &
\includegraphics[width=.3\textwidth]{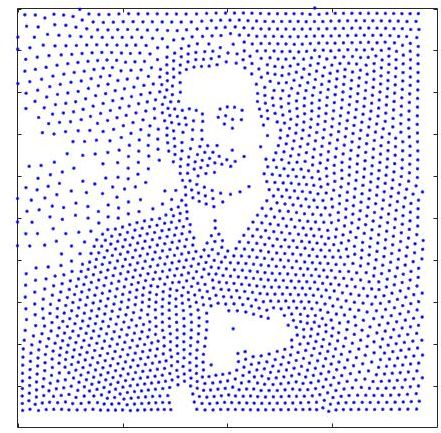} &
\includegraphics[width=.3\textwidth]{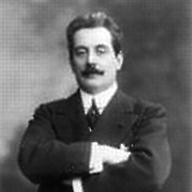}  \\
(a) & (b) & (c)
\end{tabular}
\caption{Input image (a), variables (i.e., RBF centres) (b), and reconstructed image (c).\label{FIG:EX1}}
\end{figure*}
\paragraph{Comparison between SVD methods}
We compare five SVD methods~\cite{hernandez2005slepc} on two different matrices in terms of execution time and scalability. We search for the complete set of singular values of a dense matrix and a subset of 500 singular values of a sparse matrix. All the SVD methods have a convergence tolerance of~$10^{-6}$. Table~\ref{TAB:COMPARISON2} shows that \emph{Cross} has the best results on sparse matrix, while \emph{Lanczos} and \emph{thick-restart Lanczos} have slightly worse results. On dense matrix, \emph{thick-restart Lanczos} has better results than the other methods but worse scalability properties. \emph{Cross} has the best results on 32 processes. After this preliminary test, we select \emph{Cross} as the SVD solver. We mention that a complete comparison among SVD solvers should consider additional metrics and parameters, e.g., the accuracy when computing the first singular value or the complete set of singular values.
\begin{figure*}[t]
\centering
\begin{tabular}{ccc}
\includegraphics[width=.3\textwidth]{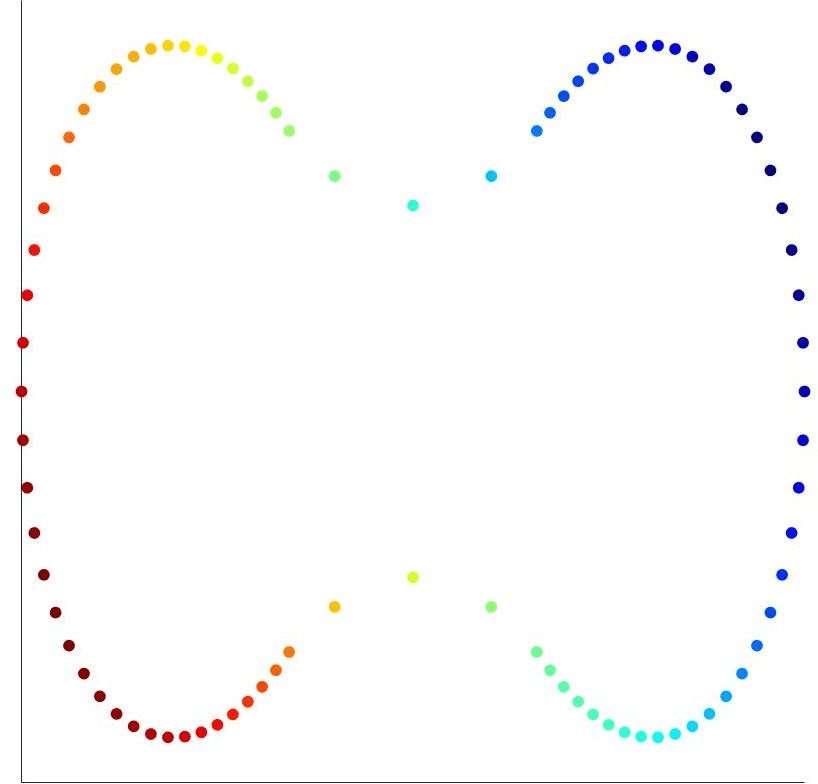} &
\includegraphics[width=.3\textwidth]{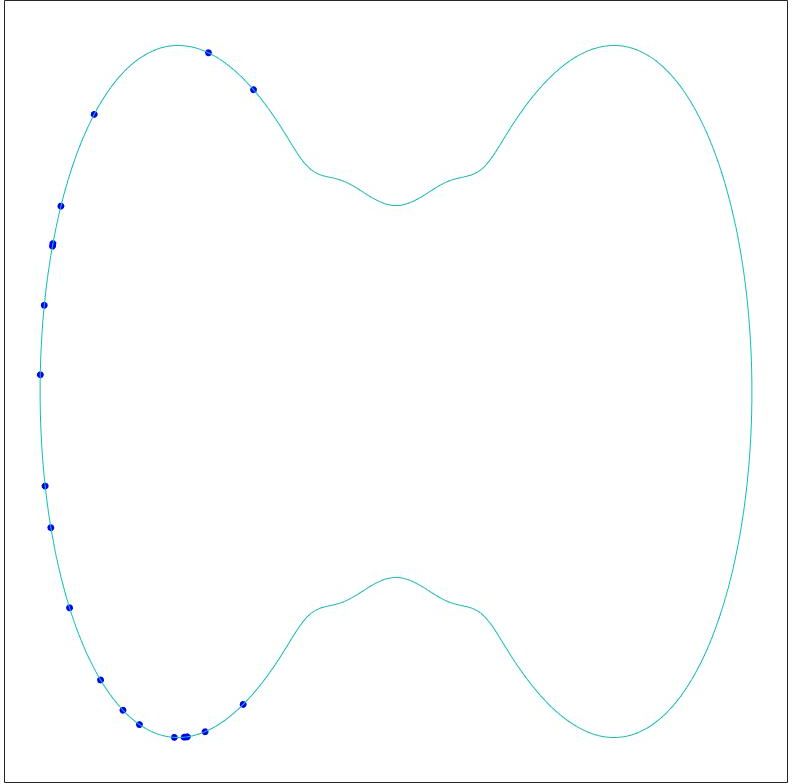} &
\includegraphics[width=.3\textwidth]{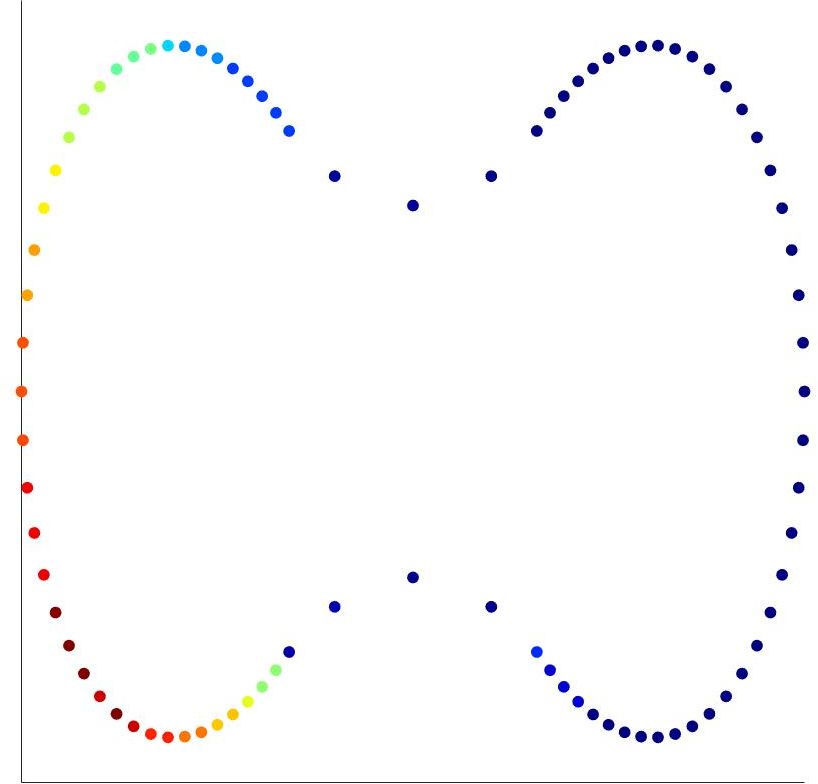}  \\
(a) & (b) & (c)
\end{tabular}
\caption{Input signal on a curve (a), variables (i.e., RBF centres, blue dots) with respect to the interpolating curve (cyan) (b), and reconstructed signal (c).\label{FIG:EX2}}
\end{figure*}
\paragraph{Scalability of functional computation}
Fig.~\ref{FIG:SCALABILITY2} and Table~\ref{TAB:SCALABILITY2} shows the scalability of the Algorithm~\ref{ALG:FUNCTIONAL2} on a dense~$5616 \times 3744$ matrix with 300 singular values; the eigenvectors matrices are dense~$5610 \times 300$ and ~$3744 \times 300$, and the singular values matrix is diagonal sparse~$300 \times 300$. The SVD passes from 24.5 seconds with 1 process to 1.4 seconds with 32 processes; the matrix-matrix multiplication passes from 4.5 seconds with 1 process to 0.4 seconds with 32 processes. The total time varies from 29 seconds with 1 process to less than 2 seconds with 32 processes; after this number of processes, both SVD and matrix-matrix multiplication increases the execution time. The efficiency with 32 processes is~$46\%$.
\begin{figure*}[t]
\centering
\begin{tabular}{ccc}
\includegraphics[width=.3\textwidth]{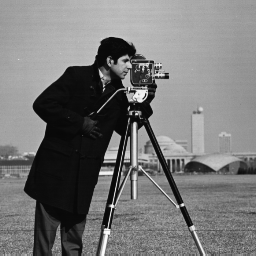} &
\includegraphics[width=.3\textwidth]{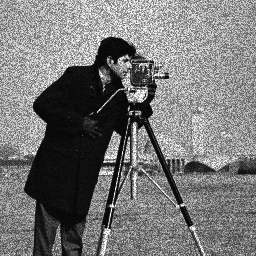} &
\includegraphics[width=.3\textwidth]{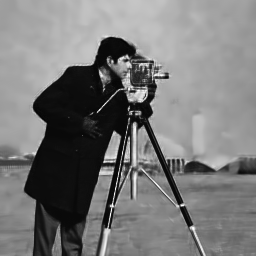}\\
(a) & (b) & (c)
\end{tabular}
\caption{Input (a), speckle noised (b), and denoised image (c).\label{FIG:EX3}}
\end{figure*}
\section{Conclusions and future work\label{SEC:CONCLUSIONS}}
We have analysed the solution of two minimisation problems of signal processing with HPC tools: approximation and denoising. For each problem, we have analysed the characteristics and results of the minimisation methods in terms of convergence and execution time. PRAXIS has shown the best results on bound-constrained problems, while ISRES has shown the best results on constrained problems. Also, we have discussed the computation of the functional and the scalability properties of the algebraic operations, including the solution of a linear system and the singular values decomposition. The two problems apply the main algebraic operations common to most signal minimisation problems; our general analysis can be extended to other signal processing problems. We show some examples of the applications: the approximation of a signal on a regular grid (Fig.~\ref{FIG:EX1}), on a curve (Fig.~\ref{FIG:EX2}), and the image denoising  (Fig.~\ref{FIG:EX3}). In future work, we want to extend the analysis to other classes of problems in signal processing (e.g., clustering) and perform the experimental tests on the novel Leonardo cluster of Cineca.\\

{\small{\paragraph{\textbf{Acknowledgements}} 
This work has been partially supported by the European Commission, NextGenerationEU, Missione 4 Componente 2, ``\emph{Dalla ricerca all’impresa}'', Innovation Ecosystem RAISE ``\emph{Robotics and AI for Socio-economic Empowerment}'', ECS00000035. Tests on CINECA Cluster are supported by the ISCRA-C project US-SAMP, HP10CXLQ1S.}}
%
\bibliographystyle{alpha}
\bibliography{refs}
%
\begin{description}
\item[\textbf{Simone Cammarasana}] is researcher at CNR-IMATI. He obtained a PhD in Computer Science at the University of Genova-DIBRIS, a post-lauream Master in Scientific Computing at the University of Sapienza-Roma, and a Master's degree in Engineering at the University of Pisa. His research interests include signals analysis, optimisation problems, and medical images.

\item[\textbf{Giuseppe Patan\'e}] is senior researcher at CNR-IMATI. Since 2001, his research is mainly focused on Computer Graphics and Shape Modelling. He is the author of scientific publications in international journals and conference proceedings, and a tutor of PhD and Post.Doc students. He is responsible for R$\&$D activities in national and European projects.
\end{description}
\end{document}